\documentclass[nofootinbib,floats,letterpaper,floatfix,eqsecnum,prd,aps,10pt]{revtex4}
\usepackage{dcolumn,epsfig,minitoc}
\usepackage{hyperref}
\usepackage{amssymb,amsmath}
\usepackage[usenames]{color}

\newcommand{\beq}{\begin{equation}}
\newcommand{\eeq}{\end{equation}}
\newcommand{\bea}{\begin{eqnarray}}
\newcommand{\eea}{\end{eqnarray}}

\begin{document}

\title{ Exact solutions with AdS asymptotics of Einstein and
Einstein-Maxwell gravity minimally coupled to a scalar field }
\author{Mariano Cadoni}
\affiliation{Dipartimento di Fisica, Universit\`a di Cagliari and INFN, Sezione di
Cagliari - Cittadella Universitaria, 09042 Monserrato, Italy. }
\author{Salvatore Mignemi}
\affiliation{Dipartimento di Matematica, Universit\`a di Cagliari and INFN, Sezione di
Cagliari - viale Merello 92, 09123 Cagliari, Italy. }
\author{Matteo Serra}
\affiliation{Dipartimento di Fisica, Universit\`a di Cagliari and INFN, Sezione di
Cagliari - Cittadella Universitaria, 09042 Monserrato, Italy. }
\date{\today}

\begin{abstract}
We propose a general method for solving exactly the static field equations
of Einstein and Einstein-Maxwell gravity minimally coupled to a scalar
field. Our method starts from an ansatz for the scalar field profile, and
determines, together with the metric functions, the corresponding form of
the scalar self-interaction potential. Using this method we prove a new
no-hair theorem about the existence of hairy black-hole and black-brane
solutions and derive broad classes of static solutions with radial symmetry
of the theory, which may play an important role in applications of the
AdS/CFT correspondence to condensed matter and strongly coupled QFTs. These
solutions include: 1) four- or generic $(d+2)$-dimensional solutions with
planar, spherical or hyperbolic horizon topology; 2) solutions with AdS,
domain wall and Lifshitz asymptotics; 3) solutions interpolating between an
AdS spacetime in the asymptotic region and a domain wall or conformal
Lifshitz spacetime in the near-horizon region.
\end{abstract}

\maketitle
\tableofcontents





\setcounter{tocdepth}{1}


\section{INTRODUCTION}

In recent years there has been a renewed growing interest for the static
black hole solutions of Einstein (and Einstein-Maxwell) gravity coupled to
scalar fields, the search for these black holes and black branes being
obviously focused on solutions dressed with a nontrivial profile of the
scalar field (scalar hair).

In the past, the interest for these solutions was basically motivated either
by the issue of the uniqueness of the Schwarzschild black hole and related
no-hair theorems \cite{Israel:1967wq,Bekenstein:1995un} or by the quest for
new black hole solutions in low-energy string models \cite%
{Gibbons:1987ps,Garfinkle:1990qj,Cadoni:1993yt,Monni:1995vu,Mayo:1996mv}.

The recent advances in string theory and in particular the AdS/CFT
correspondence \cite{Maldacena:1997re}, have shifted the focus from the
search of asymptotically flat to asymptotically anti-de Sitter (AdS)
solutions. In this context, the applications of the AdS/CFT correspondence
to holographic strongly coupled quantum field theories (QFTs) have generated
a new interesting field of application for these solutions \cite%
{Horowitz:2006ct,
Sachdev:2008ba,Hartnoll:2008vx,Hartnoll:2008kx,Herzog:2009xv,Hartnoll:2009sz}%
.

The shift to asymptotically AdS solutions permits to circumvent standard
no-hair theorems, which relate the existence of black hole solutions with
scalar hair to the violation of the positive energy theorem (PET) \cite%
{Torii:2001pg,Hertog:2006rr}. Differently from the flat case, a scalar field
in the AdS spacetime may have negative squared-mass $m^{2}$, without
destabilizing the AdS vacuum, provided $m^{2}$ is above the
Breitenlohner-Freedman (BF) bound \cite{Breitenlohner:1982bm}.

On the other hand, static black hole and black brane solutions with non
trivial scalar hair and AdS asymptotics play a crucial role in the
holographic approach to strongly coupled QFTs based on the AdS/CFT
correspondence. When the classical approximation for the bulk gravity theory
is reliable, one can deal with strongly coupled QFTs at finite temperature
in $d+1$ dimensions by investigating black holes in $d+2$ dimensions. In
this context the nontrivial, coordinate-dependent scalar hair of the black
hole solutions is interpreted either as a running coupling constant or as a
scalar condensate in the dual QFT. In the first case the bulk scalar
dynamics is very useful for holographic renormalization methods \cite%
{Skenderis:2002wp}. In the second case the scalar condensate is the origin
of a rich phenomenology in the the dual QFT, reminiscent of well-known
condensed matter systems \cite%
{Hartnoll:2008vx,Hartnoll:2008kx,Horowitz:2008bn,Herzog:2009xv,Hartnoll:2009sz, 
Cadoni:2009xm,Horowitz:2010gk,Cadoni:2011kv}. 
Following this line of reasoning, the existence of black hole solutions
with scalar hair of Einstein gravity has become a central issue in several
recent developments of the gravity/gauge field theory correspondence.

The best-known example is represented by the holographic superconductors.
Below a critical temperature the bulk gravity theory, Einstein-Maxwell with
a covariantly coupled scalar, allows for a black hole solutions with scalar
hair \cite{Hartnoll:2008vx,Horowitz:2008bn,Hartnoll:2009sz,Horowitz:2010gk}.
This corresponds to the formation of a charged condensate in the dual theory
that breaks spontaneously a global $U(1)$ symmetry. The new phase is
therefore characterized by phenomena typical of superfluid or
superconducting systems. This basic structure has been generalized to a
number of cases including, among others, Yang-Mill theories \cite%
{Ammon:2008fc} and nonminimal couplings between the scalar and the
electromagnetic (EM) field \cite%
{Charmousis:2009xr,Cadoni:2009xm,Goldstein:2009cv,
Chen:2010kn,Goldstein:2010aw,Liu:2010ka,Cadoni:2011kv}.

Another interesting feature of static black holes with scalar hair and AdS
asymptotics, which is common to several models, is the presence in the
near-horizon regime of Lifshitz-like solutions \cite%
{Cadoni:2009xm,Charmousis:2009xr,Goldstein:2009cv,Perlmutter:2010qu}. These
solutions represent a generalization of the Lifshitz spacetime,
characterized by an anisotropic scaling isometry that may be relevant for
quantum phase transitions \cite%
{Kachru:2008yh,Bertoldi:2009dt,Bertoldi:2009vn,Dehghani:2011tx,Bertoldi:2011zr,
Charmousis:2010zz,Gouteraux:2011ce}%
, in presence of a scalar field.

Despite the growing importance played by static black hole solutions with
scalar hair and AdS asymptotics, very little progress has been achieved for
what concerns the derivation of exact analytical solutions. In fact, most of
the usual methods used for finding exact, asymptotically flat, hairy
solutions \cite{Gibbons:1987ps,Garfinkle:1990qj,Cadoni:1993yt,Monni:1995vu}
do not work in AdS spacetime. Most solutions of these kind, which have been
derived and used in the literature, are numerical \cite%
{Hartnoll:2008vx,Horowitz:2008bn,Hartnoll:2009sz,Horowitz:2010gk,Cadoni:2009xm, Cadoni:2011kv}%
. Very few exact analytical solutions are known: basically we have only the
family of four-charge black holes in $\mathcal{N}=8$ four-dimensional gauged
supergravity \cite{Duff:1999gh,Lu:2009gj}, the solution with hyperbolic
horizon of Ref. \cite{Martinez:2004nb} and a few other examples some 
of them  generated
from asymptotically flat solutions 
\cite{Cai:1996eg,Cai:1997ii,Cai:2004iy,Gao:2004tu,Mignemi:2009ui}.

Obviously, this situation has a negative impact on further developments of
the subject. This is particularly true because the known no-hair theorems 
\cite{Torii:2001pg,Hertog:2006rr} put loose constraints on the existence of
black hole solutions with scalar hair. Therefore, they do not give stringent
indications that can be used when searching for exact or numerical solutions
of a given model.

Starting from these consideration, in this paper we first propose a general
method for solving the field equations of Einstein and Einstein-Maxwell
gravity minimally coupled to a scalar field $\phi$ in the static, radially
symmetric, case. Our main idea is to reverse the usual method for solving
the field equations. Usually, one determines the metric functions, the
scalar field and the EM field, for a \textsl{given} form of the
self-interaction potential $V(\phi)$. Instead of solving the field equations
for a given potential, we will assume a given profile $\phi(r)$ for the
scalar field and then we will solve the system for the metric functions and
the potential.

This method is particularly suitable for application to the AdS/CFT
correspondence. In this case the actual exact form of the potential $V(\phi)$
is not particularly relevant. What is more important is the behaviour of the
scalar field, and in particular its fall-off behaviour at $r=\infty$.

We will apply our solving method to two different but related issues. First,
we will apply it to find exact analytic solutions of Einstein and
Einstein-Maxwell gravity with scalar hair. For the scalar field we use
profiles which are very common for hairy black hole solutions in flat space,
gauged supergravity, and Lifshitz spacetime, namely harmonic and
logarithmic functions. This allows us to find exact solutions in several
situations: four or generic $d+2$ spacetime dimensions, different topologies
of the transverse space (planar, spherical, hyperbolic) and different
asymptotics (anti-de Sitter, domain wall, conformal to Lifshitz spacetime).
In particular we will derive exact solutions interpolating between an
asymptotic AdS spacetime and a near-horizon domain wall or conformal
Lifshitz spacetime. The  models that we find contain as particular case the truncation to
the abelian sector of $\mathcal{N}=8$, $D=4$ gauged supergravity.

Second, our method allows to write explicitly a formal solution of the field
equations for an arbitrary potential. This fact will be used to prove a new
no-hair theorem about the existence of black hole and black brane solutions
of Einstein and Einstein-Maxwell gravity minimally coupled to a scalar field.

The structure of the paper is as follows. In Sect.\ \ref{sect:fieldequation}
we present our general method for solving the field equations of
Einstein-Maxwell-scalar gravity in $d+2$ dimension and planar, spherical or
hyperbolic topology of the transverse sections. In Sect.\ \ref%
{sect:domainwall} we apply this method to find domain wall and conformal
Lifshitz black hole solutions in $d=2$ for the planar case. In Sect.\ \ref%
{sect:nohair} we prove a new no-hair theorem for black hole solutions 
of
Einstein-Maxwell gravity minimally coupled to a scalar field. In Sect.\ \ref%
{sect:aads} we use our method to derive planar solutions with AdS
asymptotics in $d=2$, both in the charged and uncharged case and discuss
their near-horizon behaviour. The generalization of our solutions to the $d+2
$-dimensional case and to spherical or hyperbolic solution is discussed,
respectively in Sects.\ \ref{section:ddim} and \ref{sec:spherical}. Finally
in Sect.\ \ref{sect:concl} we present our concluding remarks.

\setcounter{tocdepth}{1}


\section{ A GENERAL METHOD FOR SOLVING THE FIELD EQUATIONS}

\label{sect:fieldequation}

In this paper we will consider Einstein-Maxwell gravity in $d+2$ dimensions
(with $d\geq 2$), minimally coupled to a scalar field $\phi $, and with a
generic self-interaction potential $V(\phi )$. The action is 
\begin{equation}
A=\int d^{d+2}x\sqrt{-g}\left( \mathcal{R}-2(\partial \phi
)^{2}-F^{2}-V(\phi )\right) .  \label{action}
\end{equation}%
The ensuing field equations take the form, 
\begin{eqnarray}
&&\nabla _{\mu }F^{\mu \nu }=0\,,  \notag  \label{max_scal_eq} \\
&&\nabla ^{2}\phi =\frac{1}{4}\frac{dV(\phi )}{d\phi }\,, \\
&&\mathcal{R}_{\mu \nu }-\frac{1}{2}g_{\mu \nu }\mathcal{R}=2\left( F_{\mu
\rho }F_{\nu }^{\rho }-\frac{g_{\mu \nu }}{4}F^{\rho \sigma }F_{\rho \sigma
}\right) +2\left( \partial _{\mu }\phi \partial _{\nu }\phi -\frac{g_{\mu
\nu }}{2}\partial ^{\rho }\phi \partial _{\rho }\phi \right) -\frac{g_{\mu
\nu }}{2}V(\phi )\,.  \notag
\end{eqnarray}%
Throughout this paper we will investigate static solutions of the previous
field equations exhibiting radial symmetry. Moreover, we will consider only
purely electric solutions; magnetic solutions can be easily generated from
the electric ones using the electro-magnetic duality. We adopt a
Schwarzschild gauge to write the spacetime metric: 
\begin{equation}
ds^{2}=-U(r)dt^{2}+U^{-1}(r)dr^{2}+R^{2}(r)d\Omega _{(\varepsilon ,d)}^{2},
\label{pmetric}
\end{equation}%
where $\varepsilon =0,1,-1$ denotes, respectively, the $d$-dimensional
planar, spherical, or hyperbolic transverse space with metric $d\Omega
_{(\varepsilon ,d)}^{2}$. In these coordinates, the electric field
satisfying (\ref{max_scal_eq}) reads 
\begin{equation}
F_{tr}=\frac{Q}{R^{d}},
\end{equation}%
with $Q$ the electric charge. With the parametrization (\ref{pmetric}), the
field equations take the form 
\begin{eqnarray}
\frac{R^{\prime \prime }}{R} &=&-\frac{2}{d}(\phi ^{\prime })^{2},\quad
(UR^{d}\phi ^{\prime })^{\prime }=\frac{1}{4}R^{d}\frac{dV}{d\phi },  \notag
\label{fed} \\
(UR^{d})^{\prime \prime } &=&\varepsilon d(d-1)R^{d-2}+2\frac{d-2}{d}\frac{%
Q^{2}}{R^{d}}-\frac{d+2}{d}R^{d}V,  \notag \\
(UR^{d-1}R^{\prime })^{\prime } &=&\varepsilon (d-1)R^{d-2}-\frac{2}{d}\frac{%
Q^{2}}{R^{d}}-\frac{1}{d}R^{d}V.
\end{eqnarray}

We are mainly interested in black hole solutions of the field equations that
are asymptotically AdS. Without loss of generality we can assume that $\phi
\rightarrow 0$ as $r\rightarrow \infty $. Existence of the AdS vacuum
therefore requires $V(0)<0$ and $V^{\prime }(0)=0$. Following a widespread
convention we normalize $V(\phi )$ such that $V(0)=-d(d+1)/L^{2}$, where $L$
is the AdS length. Under this condition the simplest static black hole
solution of the field equations is given by the Schwarzschild-AdS (SAdS)
solution: 
\begin{equation}
U=\frac{r^{2}}{L^{2}}+\varepsilon -\frac{2M}{r^{d-1}},\quad R=r,\quad \phi
=0,\quad Q=0.  \label{schw}
\end{equation}

Apart from SAdS and the $Q\neq0$ Reissner-Nordstrom AdS black hole, 
the other solutions of (\ref{fed}), if they exist, will be
characterized by a non-constant profile of the scalar field $\phi$. These
solutions are very difficult to find, at least analytically. Eqs.\ (\ref{fed}%
) may be solved in closed form for some particular choice of the potential $%
V $, but for a generic potential there is no general
solving method. Moreover, it is not completely clear if and when the field
equations allow for regular black hole solutions. No-hair theorems relating
the existence of black hole solutions to the violation of the positive
energy theorem  and to the breaking of the full AdS isometry group  
have been 
discussed in the literature \cite{Torii:2001pg,Hertog:2006rr}.
Explicit solutions, analytical or numerical, are known in a few cases.
Nonetheless, a more precise statement about the existence of black hole
solutions of the field equations (\ref{fed}) is still lacking.

In this paper we will often consider solutions with scalar hair and zero
temperature (no horizon). With some abuse of terminology we will always call
these solutions ``extremal black hole solutions''. Obviously, the use of
this name is only strictly pertinent when the $T=0$ solution can be
considered as the $T\to 0$ limit of black hole solutions with a regular
horizon. As we will see in detail in the next sections, we will not be able
to show that this is the case for all the hairy zero temperature solutions
we will find. Nonetheless, we will use the word extremal black hole in the
wide sense defined above.

Usually, one solves the field equations (\ref{fed}) by determining $U$, $R$,
and $\phi$, for a \textsl{given} form of the potential $V(\phi)$. In this
paper we will approach the problem in a reversed way. Instead of solving
equations (\ref{fed}) for a given potential $V$, we will assume a given
profile $\phi(r)$ for the scalar field and then solve the system for $U(r)$, 
$R(r)$ and $V(\phi)$. Although at first sight this approach may seem rather
weird, it is very useful for at least two reasons.

First, focusing on solutions with AdS asymptotics, in particular for what
concerns application to the AdS/CFT correspondence, the actual exact form of
the potential $V(\phi)$ is not particularly relevant. What is often more
important is the behaviour of the scalar field $\phi(r)$, in particular its
fall-off behaviour at $r=\infty$. In an asymptotically AdS spacetime the $%
r\to\infty$ behaviour of a scalar field is given by 
\begin{equation}  \label{asb}
\phi\sim \frac{O_{-}}{r^{\Delta_{-}}}+\frac{O_{+}}{r^{\Delta_{+}}},\quad
\Delta_{\pm}=\frac{(d+1)\pm \sqrt{(d+1)^{2} +4m^{2}L^{2}}}{2},
\end{equation}
where $m^{2}$ is the mass of the scalar. Stability of the AdS vacuum
requires $m^{2}$ to be above the BF bound, $m^{2}\ge -(d+1)^{2}/(4L^{2})$.

Moreover, in applications of the AdS/CFT correspondence to condensed matter
physics, a nontrivial, $r$-dependent, profile of $\phi$ has a holographic
interpretation in terms of a scalar condensate in the dual QFT triggering
symmetry breaking and/or phase transitions \cite{Hartnoll:2008vx,
Horowitz:2008bn,Hartnoll:2009sz,Horowitz:2010gk,Cadoni:2009xm, Cadoni:2011kv}%
. If one is interested in reproducing phenomenological properties of
strongly-coupled condensed matter systems, the actual form of the potential $%
V$ may be rather irrelevant. Conversely, it is the behavior of the scalar
condensate that contains more physical information \cite%
{Hartnoll:2008vx,Horowitz:2008bn,Hartnoll:2009sz,Horowitz:2010gk,Cadoni:2009xm, Cadoni:2011kv}%
.

Second, our approach is very useful for setting up a new no-hair theorem
about the existence of black hole solutions of the field equations. In fact,
our method allows us to write explicitly a -- albeit formal -- solution of
the field equations for an arbitrary potential. This result will be used in
Sect. \ref{sect:nohair}, to prove a new no-hair theorem about the existence
of black hole solutions of minimally coupled Einstein-Maxwell-scalar gravity.

Our method for solving the field equations (\ref{fed}) works as follows.
Assuming that the $r$-dependence of the scalar field $\phi (r)$ is given,
and introducing the new variables $F$, $Y$ and $u$ defined as 
\begin{equation}
F(r)=-\frac{2}{d}(\phi ^{\prime })^{2},\quad R=e^{\int Y},\quad u=UR^{d},
\label{nv}
\end{equation}%
the field equations (\ref{fed}) become 
\begin{eqnarray}
&&Y^{\prime }+Y^{2}=F,\quad (u\phi ^{\prime })^{\prime }=\frac{1}{4}%
e^{d\!\int \!Y}\frac{dV}{d\phi },  \label{z1} \\
&&u^{\prime \prime }-(d+2)(uY)^{\prime }=-2\varepsilon (d-1)e^{(d-2)\int
Y}+4Q^{2}e^{-d\int Y},  \label{z2} \\
\quad u^{\prime \prime } &=&\varepsilon d(d-1)e^{(d-2)\int Y}+2\frac{d-2}{d}%
Q^{2}e^{-d\int Y}-\frac{d+2}{d}e^{d\int Y}V.  \label{z3}
\end{eqnarray}

The first equation in (\ref{z1}) is a first-order nonlinear equation for $Y$%
, known as the Riccati equation, which can be solved in a number of cases.
Once the solution for $Y$ has been found we can integrate Eq. (\ref{z2}),
which is linear in $u$, to obtain 
\begin{equation}
u=R^{d+2}\left[ \int \left( 4Q^{2}\int \frac{1}{R^{d}}-2\varepsilon
(d-1)\int R^{d-2}-C_{1}\right) \frac{1}{R^{d+2}}+C_{2}\right] ,  \label{sol1}
\end{equation}%
where $C_{1}$ and $C_{2}$ are integration constants. Finally, we can
determine the potential $V(\phi )$ by using Eq. (\ref{z3}), 
\begin{equation}
V=\frac{d^{2}(d-1)}{d+2}\frac{\varepsilon }{R^{2}}+2\frac{d-2}{d+2}\frac{%
Q^{2}}{R^{2d}}-\frac{d}{d+2}\frac{u^{\prime \prime }}{R^{d}},
\label{potential}
\end{equation}
while the metric functions read (cfr.\ \ref{nv}) 
\begin{equation}  \label{mf}
R=\Lambda e^{\int Y},\qquad U={\frac{u}{R^d}},
\end{equation}
where we have introduced an integration constant $\Lambda$ coming from the
integral of $Y$.

In the following sections we will use this method to find solutions of
minimally coupled Einstein-Maxwell-scalar gravity in different spacetime
dimensions and for planar, spherical, and hyperbolic topologies of the $d$%
-dimensional transverse section of the spacetime.


\section{DOMAIN WALLS AND SOLUTIONS CONFORMAL TO
LIFSHITZ}

\label{sect:domainwall} In this section, we consider the case of
(3+1)-dimensional spacetime, i.e.\ $d=2$, and black brane solutions, i.e. $%
\varepsilon=0$. This is the most useful case for applications to holography.
These solutions will be generalized to $d+2$ spacetime dimensions in section %
\ref{section:ddim} and to black holes with spherical ($\varepsilon=1$) or
hyperbolic ($\varepsilon=-1$) symmetry in Sect.\ \ref{sec:spherical}.

Our method for solving the field equations (\ref{fed}) requires an ansatz
for the scalar field. In this section we wish to find domain wall and
Lifshitz-like solutions. Usually, these solutions appear when the scalar
behaves as $\log r$ \cite{Charmousis:2009xr,Cadoni:2009xm,Goldstein:2009cv,
Goldstein:2010aw,Cadoni:2011kv}. The most natural ansatz is therefore: 
\begin{equation}  \label{gamma}
\gamma\phi=\log \frac{r}{r_{-}},
\end{equation}
where $\gamma$ and $r_-$ are constants. Note that $r_-$ has no particular
physical meaning, but simply sets a length-scale.

\subsection{Uncharged (Domain wall) solutions}

\label{sect:unchargedbb}

Let us consider the solutions (\ref{sol1}), (\ref{mf}) for $d=2$, $%
\varepsilon=0$. At first we examine the simplest case in which $Q=0$.
Choosing $C_{1}=0$ and scaling the constants $C_2$ and $\Lambda$ to 1, one
gets 
\begin{equation*}
U=R^{2}.
\end{equation*}
Hence, for this choice of the parameters, the solution takes the form of a
domain wall, 
\begin{equation}  \label{e12}
ds^{2}= U(-dt^{2}+dx^{2}+dy^{2}) + U^{-1}dr^{2}.
\end{equation}
Notice that in the relevant cases, even when $C_{1}\neq 0$ the corresponding
term in Eq. (\ref{sol1}) in the $r\to \infty$ limit is subleading, and
therefore the solution is still asymptotical to a domain wall.

With the ansatz (\ref{gamma}) the Riccati equation is solved by 
\begin{equation}  \label{solution}
Y= \frac{\alpha}{r},\quad \alpha(\alpha-1)=-\frac{1}{\gamma^{2}}.
\end{equation}

Parametrizing $\alpha $ and $\gamma $ as 
\begin{equation}
\gamma ^{-1}=h\alpha =\frac{h}{h^{2}+1},  \label{param}
\end{equation}%
the solution takes the form 
\begin{equation}
U=\left( \frac{r}{r_{-}}\right) ^{\frac{2}{1+h^{2}}}-C_{1}\,\left( \frac{r}{%
r_{-}}\right) ^{-\frac{1-h^{2}}{1+h^{2}}},\qquad R=\left( \frac{r}{r_{-}}%
\right) ^{\frac{1}{1+h^{2}}},  \label{e11}
\end{equation}%
with potential 
\begin{equation}
V=-\frac{2(3-h^{2})}{(1+h^{2})^{2}r_{-}^{2}}\;e^{-2h\phi }.  \label{pot}
\end{equation}%
Hence the potential has a simple exponential form. If in the theory a
length-scale $L$ is present, as happens for instance when the exponential
potential arises as near-horizon approximation of an asymptotically AdS
spacetime, one can trade $r_{-}$ for $L$ using the invariance of the field
equations under rescaling of $V\to \lambda V,U\to \lambda U$, yielding $V=-[{%
\frac{2(3-h^{2})}{(1+h^{2})L^{2}}}\,]e^{-2h\phi }$.

In the extremal case, $C_{1}=0$, the solution (\ref{e11}) has the typical
form of a single-scalar domain wall solution, $ds^{2}=(Ar)^{\delta }(\eta
_{\mu \nu }dx^{\mu }dx^{\nu })+(Ar)^{-\delta }dr^{2}$ \cite%
{Kaitscheider:2009as,Boonstra:1998mp,Perlmutter:2010qu}. These solutions
preserve only the Poincar\'{e} isometry of the 3D transverse space and are
related to solitonic supergravity domain-walls, resulting from various
dimensional reductions of 10 and 11-dimensional maximal supergravity
theories.

Moreover, domain wall solutions are conformal to AdS spacetime and have a
consistent holographic interpretation, in terms of a dual QFT with only
relativistic symmetry, for $\delta\ge 1$, which in our case implies $%
h^{2}\le 1$.

One can easily calculate the curvature invariants for solution (\ref{e11}), 
\begin{eqnarray*}
\mathcal{R} &=&\frac{2}{(1+h^{2})^{2}r_{-}^{2}}\left[ 3(h^{2}-2)x^{\frac{%
-2h^{2}}{1+h^{2}}}-h^{2}\mu x^{\frac{-3-h^{2}}{1+h^{2}}}\right] , \\
\mathcal{R}_{\mu \nu }\mathcal{R}^{\mu \nu } &=&\frac{4}{%
(1+h^{2})^{4}r_{-}^{4}}\left[ (3h^{4}-9h^{2}+9)x^{\frac{-4h^{2}}{1+h^{2}}%
}+3h^{2}(1-h^{2})\mu x^{-3}+h^{4}\mu ^{2}x^{\frac{-6-2h^{2}}{1+h^{2}}}\right]
,
\end{eqnarray*}%
where $x=r/r_{-}$, showing that $r=0$ is a curvature singularity.

For $h^{2}\leq 3$ and $C_{1}>0$, our solution (\ref{e11}) represents a black
brane with domain wall asymptotics, a singularity at $r=0$ and a horizon at $%
r_{h}=C_{1}^{\;(1+h^{2})/(3-h^{2})}r_{-}$. The horizon is regular and has
negative curvature ($\mathcal{R}(r_{h})<0\,$, $\mathcal{R}_{\mu \nu }%
\mathcal{R}^{\mu \nu }(r_{h})\neq 0$). The domain wall solution for $C_{1}=0$
display a naked singularity at the origin and may be seen as the $T=0$
extremal limit of (\ref{e11}). Notice however, that the solution at finite
temperature breaks the Poincar\'{e} isometry of the extremal domain wall
solution. For $h^{2}>3$, the solution is still valid, but its physical
interpretation is less clear.

\subsection{ Solutions conformal to Lifshitz spacetime}

\label{sect:chargedbb}

Let us now consider the case of nonvanishing electric charge. In this case
it is convenient to adopt the parametrization 
\begin{equation}
\alpha =\frac{h^{2}}{h^{2}+4},\qquad \gamma ^{-1}=\frac{2h}{h^{2}+4}.
\label{paramc}
\end{equation}%
Using again the ansatz (\ref{gamma}) and Eq. (\ref{solution}), the solution (%
\ref{sol1}) reads 
\begin{eqnarray}
U &=&\frac{Q^{2}}{\Lambda ^{4}}\frac{(4+h^{2})^{2}r_{-}^{2}}{%
(4-h^{2})(2-h^{2})}\;\left( \frac{r}{r_{-}}\right) ^{2\frac{4-h^{2}}{4+h^{2}}%
}\left[ 1-C_{1}\,\left( \frac{r}{r_{-}}\right) ^{-\frac{4-h^{2}}{4+h^{2}}%
}+C_{2}\,\left( \frac{r}{r_{-}}\right) ^{-4\frac{2-h^{2}}{4+h^{2}}}\right] ,
\notag  \label{e11a} \\
R &=&\Lambda \left( \frac{r}{r_{-}}\right) ^{\frac{h^{2}}{4+h^{2}}},
\end{eqnarray}%
with potential 
\begin{equation}  \label{ppot}
V=-\frac{Q^{2}}{\Lambda ^{4}}\left[\frac{4}{2-h^{2}}\;e^{-2h\phi }+ \frac{%
2h^{2}(3h^{2}-4)}{(4-h^{2})(2-h^{2})}\;C_{2}\,e^{-4\phi /h}\right],
\end{equation}%
where the integration constant $C_{1}$ and $C_{2}$ of (\ref{sol1}) have been
rescaled. The solutions hold for $h^{2}\neq 2,4$.

Setting $\Lambda^2=r_- Q$, the potential becomes independent from the
electric charge. If an extra length scale $L$ is present, one can, like in
the uncharged case, trade $r_{-}$ for $L$ in the potential (\ref{ppot}), so
that it acquires a factor $1/L^2$. The constant $C_{2}$ is a parameter of
the action, that can be chosen to vanish. In such case, one is left with an
exponential potential like in (\ref{pot}).

For $C_{2}=0$, the extremal $C_{1}=0$ case in Eq.\ (\ref{e11a}) represents a
solution which is conformal to the Lifshitz spacetime, 
\begin{equation}
ds^{2}=l^{2}\left( -\bar{r}^{2z}dt^{2}+\frac{d\bar{r}^{2}}{\bar{r}^{2}}+\bar{%
r}^{2}dx^{i}dx^{i}\right) .  \label{ls}
\end{equation}
This can be easily shown by setting $\bar r= (r/r_{-})^{-h^{2}/(4+h^{2})}$
in Eq.\ (\ref{e11a}). The metric (\ref{e11a}) is conformal (with conformal
factor $\bar r^{-4}$) to the Lifshitz metric (\ref{ls}) with $z=3-4/h^2$ and 
$l=r_{-}¥\sqrt{(4-h^{2})(2-h^{2})}/h^2$. Obviously the anisotropic scaling
transformation between space and time 
\begin{equation}  \label{lts}
t\to \lambda^{z} t,\quad \bar r\to \lambda^{-1} \bar r,\quad x^{i}\to
\lambda x^{i},
\end{equation}
which is an isometry of the Lifshitz metric (\ref{ls}), is not longer an
isometry of our solution (\ref{e11a}). However, its conformality with
Lifshitz implies that it scales with a definite weight under the anisotropic
scaling transformation (\ref{lts}): $ds^{2}\to \lambda^{4}ds^{2}$. In the
remaining of this paper we will denote solutions which are conformal to
Lifshitz spacetime simply as "conformal Lifshitz''.

For $h^2<2$, $C_{2}=0$ and $C_{1}>0$ the solution represents a black brane
with asymptotics conformal to Lifshitz, a singularity at $r=0$ and a regular
horizon at $r_h =C_1^{\;(4+h^2)/(4-h^2)}r_-$. The conformal Lifshitz
solution with $C_{1}=0$ may be seen as the $T=0$ extremal limit of (\ref%
{e11a}). The solution at finite temperature do not follow the simple scaling
behavior of the extremal solution.

If instead $C_2\neq0$, two horizons may be present, depending on the value
of the parameters. Moreover, the term in $C_2$ becomes dominant for $%
r\to\infty$ if $h^2>2$. In any case, the solutions (\ref{e11a}) constitute a
two-parameter family parametrized by $C_1$ and $Q$.

\subsection{Alternative approach}

When $C_2=0$, the solutions of this section can be obtained also by means of
a more traditional approach, introduced in \cite{Gibbons:1987ps} and
developed in several papers \cite%
{Mignemi:1988qc,Garfinkle:1990qj,Cadoni:1993yt,Monni:1995vu}.

Parametrizing the metric and the electric field as 
\begin{equation}
ds^{2}=-e^{2\nu }dt^{2}+e^{2\nu +4\rho }d\xi ^{2}+e^{2\rho
}(dx^{2}+dy^{2}),\qquad F_{t\xi }=e^{2\nu }Q,
\end{equation}%
where $\nu =\nu (\xi )$, $\rho =\rho (\xi )$ and $\phi =\phi (\xi )$, one
can in fact reduce the field equations to the form of a dynamical system,
that admits a three-parameter family of regular black brane solutions. Exact
solutions can be obtained in a special two-parameter case and coincide with
those obtained above. The third parameter is presumably related to the
scalar charge. More details will be given elsewhere \cite{m}.

\section{ A NO-HAIR THEOREM}

\label{sect:nohair}

In the previous section we have seen that minimally coupled
Einstein-Maxwell-scalar gravity in $3+1$ dimensions allows for black brane
solutions with either domain wall or conformal Lifshitz asymptotics. All
these solutions correspond to an exponential potential $V(\phi)$ in the
uncharged (domain wall) and charged (conformal Lifshitz) case, respectively.
We will see later in Sects. \ref{section:ddim} and \ref{sec:spherical}, that
these results hold also in $d+2$ dimensions and also for black holes, i.e.\
solutions with spherical horizons.

An important issue in this context is the question about the existence of
regular, static black hole solutions of Einstein-scalar gravity with AdS
asymptotics beyond the Schwarzschild and Reissner-Nordstrom solutions, i.e.\ of
solutions endowed with non trivial scalar hair.

This problem has been already discussed in the literature. In particular, it
has been argued that  necessary conditions for the existence of such black
hole solutions are the violation of the PET and, in theories where 
the scalar field has negative local maxima, the breaking of the full AdS 
isometry  group \cite{Torii:2001pg,Hertog:2006rr}.
These results rule out black hole solutions with scalar hair and positive
squared-mass $m^{2}$,\footnote{%
As a consequence, asymptotically flat black holes are ruled out completely
because the Minkowski vacuum would be unstable.} but allows for the
existence of these solutions when $m^{2}$ is negative but above the BF
bound. Some examples of regular (analytical or numerical) AdS black hole
solutions with scalar hair are known. An analytical solution has been
obtained in Ref. \cite{Martinez:2004nb}, whereas numerical solutions have
been found in the context of designer gravity \cite{Hertog:2004dr},
Einstein-Maxwell gravity with covariantly coupled scalar \cite%
{Hartnoll:2008vx,Horowitz:2008bn,Hartnoll:2009sz,Horowitz:2010gk} and
Einstein-Maxwell dilaton gravity \cite{Cadoni:2009xm,Cadoni:2011kv}.

We prove here, using the reformulation of the field equations discussed in
Sect.\ \ref{sect:fieldequation}, a new no-hair theorem about the existence
of regular hairy black hole solutions of Einstein-Maxwell-scalar gravity. We
will consider for simplicity the $d=2$ case, but our theorem can be
trivially generalized to arbitrary $d+2$ dimensions. We will first consider
the uncharged and planar case $Q=\varepsilon=0$ and then we will generalize
our argument to the charged and $\varepsilon=\pm 1$ cases.

A key ingredient for our argument is the existence of an extremal $T=0$
hairy black hole solution. We will prove the validity of the following three
statements about black hole solutions of Einstein gravity minimally coupled
to a scalar field:\newline
\textsl{$1)$ One-parameter families of asymptotically AdS black brane
solution with nontrivial scalar hair exist only if the field equations (\ref%
{fed}) admit an extremal $T=0$, $U=R^{2}$ solution.\newline
$2)$ Black-brane solutions that asymptotically approach the domain wall
solution (\ref{e11}) exist in some range of the parameters for the case of
an exponential potential $V(\phi )$.\newline
$3)$ The allowed asymptotically AdS hairy black brane solutions necessarily
have a scalar hair that depends on the black brane temperature $T$.
Solutions with temperature-independent scalar hair exist only for the case
of domain wall spacetimes (\ref{e11}).}

In order to prove part $1)$ of the theorem we start from equation (\ref{sol1}%
), set $d=2,\, \varepsilon=0$ and fix the physically irrelevant integration
constant $C_{2}=1$. We get the general form of the solution of the field
equation (\ref{fed}) for the metric function $U$ 
\begin{equation}  \label{e20}
U= R^{2}\left(1- C_{1} \int \frac{1}{R^{4}} \right).
\end{equation}
The integration constant $C_{1}$ is determined in terms of the mass $M$ (or
equivalently of the temperature $T$) of the solution.

Assuming the existence of a one-parameter family of black brane solutions
with a regular horizon at $r=r_{h}$ we have to require $U(r_{h})=0$, $%
R(r_{h})\neq 0$. This implies that the horizon is determined by the equation 
\begin{equation}  \label{e21}
1- C_{1} \int \frac{1}{R^{4}}=0,
\end{equation}
whereas for $C_{1}=0$ we have an extremal $T=0$ domain wall solution with $%
U=R^{2}$.

Let us now assume that the field equations do not admit the $C_{1}=0$
extremal solution. Inserting Eq. (\ref{e20}) into the third field equation
in (\ref{fed}), $C_{1}$ becomes completely determined in terms of the
functions $R$ and $\phi$. This, at least in principle, can be used to
eliminate the integration constant $C_{1}$ from the field equations (\ref%
{fed}), which can now be used to determine the solutions for $R$ and $\phi$.
As a consequence, the solutions for $R$ and $\phi$ will not depend on $C_{1}$%
, i.e. they will be \textsl{temperature independent}.

Let us now pick up a particular --albeit generic-- solution of the field Eq.
(\ref{fed}) with $C_{1}=C_{1}^{(0)}$, denote it with $(U_{0},\,R_{0},\,\phi
_{0})$ and decompose the general solution of the field equations (\ref{fed})
as follows: 
\begin{equation}
(U=U_{0}+\tilde{U}(C_{1},r),\quad \phi ,\text{ \ }R).  \label{e23}
\end{equation}%
Because $R$ and $\phi $ do not depend on $C_{1}$, we must have $\phi =\phi
_{0}$ and $R=R_{0}$. Substitution of Eq. (\ref{e23}) into Eq. (\ref{fed})
gives $(\tilde{U}R^{2})^{\prime \prime }=0,\,(\tilde{U}R^{2}\phi ^{\prime
})^{\prime }=0$, which implies $\phi =c\log \frac{r}{r_{0}}$, with $c,r_{0}$
integration constants. According to Eqs.\ (\ref{gamma})-(\ref{pot}), this is
only possible for an exponential potential $V(\phi )$ and gives the domain
wall solution (\ref{e11}). From this, part $1)$ and part $2)$ of the no-hair
theorem follow immediately.

Obviously, if the field equation allow the $C_{1}=0$ solution the previous
derivation fails. We can choose in Eq. (\ref{e23}) $U_{0}$ as the $C_{1}=0$
solution, whereas $\phi$ and $R$ do not need to be independent from $C_{1}$.

Statement $3)$ can be proved with a slight modification of the previous
argument. One begins by noticing that, owing to the first equation in (\ref%
{fed}), a temperature-independent scalar hair implies that also the function 
$R$ is temperature-independent. One then assumes the existence of a
one-parameter family of black brane solutions of the field equations (\ref%
{fed}) $(U(C_{1}), \phi, R)$ with $\phi$ and $R$ independent of $C_{1}$.
Repeating the argument starting from Eq.\ (\ref{e23}) one easily finds that
the one-parameter family of hairy solutions $(U(C_{1}), \phi, R)$ exists
only in the case of an exponential potential and is given by the black brane
solutions (\ref{e11}).

The previous derivation can be easily extended to the charged case $Q\neq 0$
and to $\varepsilon=\pm 1$. The only new ingredient is that now the general
solution of the field equations (\ref{fed}) is determined by 
\begin{equation}  \label{nh}
U=R^{2}\left[1+\int \left( 4Q^{2}\int \frac{1}{R^{2}}-2\varepsilon
r-C_{1}\right) \frac{1}{R^{4}}\right],
\end{equation}%
rather than by Eq.\ (\ref{e20}). Since for $Q\neq 0$ or $\varepsilon\neq0$
the solution (\ref{nh}) with $C_{1}=0$ is no longer given by $U=R^{2}$, in
general, it will not necessarily be an extremal $T=0$ solution. This is
related to the fact that $C_{1}$ will now be determined not only in terms of
the black hole mass $M$, but in terms of $Q$ as well. As a consequence, the
spacetime will in general have an inner and outer horizon. Statement $1)$
then holds in a much weaker form: One-parameter families of asymptotically
charged AdS black hole solution with nontrivial scalar hair exist only if
the field equations (\ref{fed}) admit a black hole solution with $C_{1}=0$.
Because the $C_{1}=0$ solution is not necessarily extremal this statement is
not particularly useful.

On the other hand statements $2)$ and $3)$ do not depend on the existence of
an extremal solution. Their generalization to the charged and $%
\varepsilon=\pm 1 $ case is almost trivial. Statement $2)$ now affirms that
charged black brane/black hole solutions that asymptotically approach the
conformal Lifshitz spacetime (\ref{e11a}) exist, in some range of the
parameters, for the case of an exponential potential $V(\phi)$. Statement $%
3) $ still remains true in the form given above also for the case of charged
and $\varepsilon=\pm1$ black holes.

Concerning statement $3)$ it is important to stress that this theorem does
not apply to the case of a nonminimal coupling between the scalar and the
gauge field. In the latter case we have an additional term depending on the
derivative of the coupling function between the scalar and $F^{2}$ in the
last equation of (\ref{fed}). The effect of this term is to allow for
solutions with two integration constants $r_{\pm}$, with $Q\sim r_{-}r_{+}$, 
$T\sim (r_{+}-r_{-})$, whereas the scalar field depends on $r_{-}$ only.
Hence the scalar hair is independent from the black hole temperature but is
related to $Q$. This result is perfectly consistent with the
well-established existence of black holes with temperature-independent
scalar charges in models with a non-minimally coupled scalar field.

We conclude this section by listing the classes of static black hole
solutions of Einstein-Maxwell-scalar gravity that may exist in view of the
above no-hair theorems.

\begin{itemize}
\item Models with an exponential potential admit, at least in some range of
the parameters, a one-parameter family of domain wall ($Q=0,\varepsilon=0$),
conformal Lifshitz ($Q\neq 0,\varepsilon=0$), black brane or black hole ($%
Q=0 $ or $Q\neq 0$ and $\varepsilon=\pm 1$) solutions.

\item The existence of asymptotically AdS uncharged black brane solutions
with scalar hair is tightly constrained. Apart from the violation of the
PET, a further necessary, but not sufficient, condition for their existence
is that the field equations allow for an extremal $T=0$ solution.

\item The existence of hairy asymptotically AdS charged black branes, or
charged and uncharged black holes, is very loosely constrained by the above
no-hair theorem.

\item For all cases (charged and uncharged, black branes and black holes)
the allowed hairy AdS  solutions must have a temperature-dependent
scalar hair.
\end{itemize}

\section{ASYMPTOTICALLY ADS SOLUTIONS}

\label{sect:aads}

In this section we will derive asymptotically AdS solutions with scalar hair
of the field equations (\ref{fed}) for $d=2$ and $\varepsilon=0$. As
explained in Sect.\ \ref{sect:fieldequation}, in order to have
asymptotically AdS solution we require the potential to satisfy, $%
V(0)<0,\,V^{\prime}(0)=0$ and we normalize $V$ using $V(0)=-6/L^{2}$.

As usual, the starting point of our solving method is an ansatz for the
scalar field. Inspired by known solutions in flat spacetime and in gauged
supergravity \cite%
{Gibbons:1987ps,Garfinkle:1990qj,Cadoni:1993yt,Monni:1995vu,Duff:1999gh}, we
use an ansatz in which $\phi$ is expressed in terms of a four-dimensional
harmonic function $X$, 
\begin{equation}  \label{harmf}
\gamma \phi=\log X, \qquad X=1-\frac{r_{-}}{r},
\end{equation}
where $\gamma$ and $r_{-}$ are constants. Notice that with this ansatz, in
the asymptotical AdS region the scalar field $\phi$ is a tachyonic
excitation with mass above the BF bound in 4D, $m^{2}= -2/L^{2}$. Expanding
Eq. (\ref{harmf}) near $r=\infty$ and comparing with Eq.\ (\ref{asb}) one
finds that the asymptotic behaviour of the scalar field is characterized by $%
\Delta_{\pm}=2,1$ and by $O_{+}= r_{-}\, O_{-}/2$. This tells us that we are
dealing with so called designer gravity models \cite{Hertog:2004dr}.

Given the ansatz (\ref{harmf}) the Riccati equation (\ref{z1}) can be solved
in terms of the harmonic function $X$ to give, 
\begin{equation}  \label{f2}
R=\Lambda rX^{\beta+\frac 1 2},\qquad \beta^2-\frac 1 4=-\frac{1}{\gamma^{2}}%
.
\end{equation}
where $\Lambda$ can be set to 1 without loss of generality, if $\varepsilon=0
$ and $Q=0$. Notice that the previous equation implies 
\begin{equation}  \label{g3}
-\frac{1}{2}<\beta<\frac{1}{2}.
\end{equation}
As usual we will proceed by discussing separately uncharged and charged
solutions.

\subsection{Uncharged solution}

\label{sect:uncharged}

Let us set $Q=0$, in Eq.\ (\ref{sol1}) and first consider the $C_{1}=0$
extremal solutions. The constant $C_{2}$ essentially determines the
normalization of the potential. This is fixed by choosing $C_{2}=1/L^2$.
With these assumptions, Eq.\ (\ref{sol1}) and (\ref{potential}) give,
respectively, the solution for the metric and the scalar potential 
\begin{eqnarray}
U &=&R^{2}=\frac{r^{2}}{L^{2}}\left( 1-\frac{r_{-}}{r}\right) ^{2\beta +1},
\label{f3} \\
V_{1}(\gamma ,\phi ) &=&-\frac{2}{L^{2}}e^{2\gamma \beta \phi }\left[
2-8\beta ^{2}+(1+8\beta ^{2})\cosh (\gamma \phi )-6\beta \sinh (\gamma \phi )%
\right] .  \label{pot1}
\end{eqnarray}%
One can easily check that these solutions represent domain walls with AdS$%
_{4}$ asymptotics. Calculating the periodicity of the 2D Euclidean section
one can also check that the solution is an extremal $T=0$ solution.

The potential (\ref{pot1}) interpolates smoothly between the asymptotic AdS
region at $\phi =0$ and a $\phi \rightarrow \infty $ region (a near-horizon
region) where the potential behaves exponentially, 
\begin{equation}
V(\phi )=-\frac{(2\beta +1)(4\beta +1)}{L^{2}}\ e^{\gamma (2\beta -1)\phi }.
\label{pot1as}
\end{equation}%
Moreover, it contains as a special case, $\beta =0$, the potential resulting
from truncation to the abelian sector of $\mathcal{N}=8$, $D=4$ gauged
supergravity \cite{Hertog:2004dr}, 
\begin{equation}
V(\phi )=-\frac{2}{L^{2}}(\cosh 2\phi +2).  \label{gaugedsu}
\end{equation}%
In this case the solution (\ref{f3}) takes the particularly simple form 
\begin{equation}
U=R^{2}=\frac{r^{2}}{L^{2}}-\frac{rr_{-}}{L^{2}},\,\,\phi =\frac{1}{2}\log
\left( 1-\frac{r_{-}}{r}\right) .  \label{f4b}
\end{equation}

The model described by the potential (\ref{pot1}) becomes very simple also
for $\beta=1/4$, 
\begin{equation}  \label{f5}
V(\phi)=-\frac {6} {L^{2}}\cosh\frac{2\phi}{\sqrt 3}.
\end{equation}

The potential (\ref{pot1}) remains invariant under the duality
transformation 
\begin{equation}  \label{duality}
\phi\to-\phi,\quad \beta\to -\beta.
\end{equation}
This symmetry of the action can be used to generate a new dual solution from
Eq. (\ref{f3}), 
\begin{equation}  \label{k1}
U=R^{2}=\frac {r^{2}} {L^{2}} X^{-2\beta+1},\quad \gamma \phi= -\log X.
\end{equation}
Notice that for the supergravity model (\ref{gaugedsu}) the symmetry
transformation is simply $\phi\to-\phi$, whereas solution (\ref{f4b})
becomes self-dual.

Because the model with potential (\ref{pot1}) admits the $C_{1}=0$ solution,
statement $1)$ of the no-hair theorem discussed in the previous section
implies that also non extremal black brane solutions with $C_{1}\neq 0$ can
in principle exist. Unfortunately, our method does not allow to find such a
solutions. Naively, one could think that these solutions can be derived just
by using Eq.\ (\ref{f2}) into Eq.\ (\ref{sol1}) with $C_{1}\neq 0$. This is
not the case not only because the resulting potential $\tilde V$ is
different from (\ref{pot1}) but, more importantly, because $\tilde V$ will
depend explicitly on $C_{1}$, which instead should be a free integration
constant related to the mass of the solution.

Notice that a one-parameter family of solutions can be generated from Eq.\ (%
\ref{sol1}) with $C_{1}\ne 0$ by using the invariance of the field equations
under the rescaling $R\to \lambda R$, to let the potential depend only on
the ratio $C_{2}/C_{1}$, whereas $U$ depends on both $C_{2}$ and $C_{1}$.
However, in this case the solution $C_{1}=0$ is not allowed. Hence the
no-hair theorem of the previous section implies that this family of
solutions are not black branes. Thus non-extremal black brane solutions of
models with the potential (\ref{pot1}), if they exist, have to be found
numerically.

The hairy extremal solution (\ref{f3}) interpolates between an AdS vacuum at 
$r=\infty$ and a domain wall solution (\ref{e11}) near $r=r_{-}$. This can
be easily shown by expressing solution (\ref{f3}) in the near-horizon
approximation $r\sim r_{-}$. Shifting $r\to r+r_{-}$ and expanding near $r=0$
one finds at leading order, 
\begin{equation}  \label{f6}
\gamma\phi= \ln \frac{r}{r_{-}},\qquad U=R^{2}=A^{2}\left(\frac{r}{r_{-}}%
\right)^{2\beta +1},\qquad A=\frac{r_{-}}{L}.
\end{equation}
As expected, this solution is easily recognized, just by setting $%
\beta=(1-h^{2})/(2+2h^{2})$ and by rescaling $V$ (in the way explained after
Eq. (\ref{e11})) as the exact solutions (\ref{e11}) with $C_{1}=0$ of a model
with near-horizon exponential potential (\ref{pot1as}).

The near-extremal solutions with a horizon, corresponding to solutions (\ref%
{e11}) with $C_{1}\neq 0$ are given by 
\begin{equation}  \label{f6r}
U=A^{2}\left(\frac{r}{r_{-}}\right)^{2\beta+1}-\mu \left(\frac{r}{r_{-}}%
\right)^{-2\beta} ,
\end{equation}
whereas $\phi$ and $R$ are given as in Eq.\ (\ref{f6}). We stress again that
Eq. (\ref{f6}) and (\ref{f6r}) are exact solution of the near-horizon
approximate form of the potential (\ref{pot1as}) but only leading-order
solution of the near-horizon approximation for the exact potential (\ref%
{pot1}).

The near-horizon, extremal and near-extremal solution, corresponding to the
dual solution (\ref{k1}) can be easily obtained from Eq. (\ref{f6r}) just by
using the duality transformation (\ref{duality}).

\subsection{Charged solutions}

Following the same steps described in the previous subsection we now derive $%
Q\neq 0$ hairy black brane solutions of the field equations (\ref{fed}) with
AdS asymptotics.

The $C_{1}=C_{2}=0$ solution, corresponding to the ansatz (\ref{harmf}) for
the scalar, is obtained by substituting the solution (\ref{f2}) of the
Riccati equation  into Eq. (\ref{sol1}),

\begin{eqnarray}
U &=&\frac{r^{2}}{L^{2}}\left( 1-\frac{r_{-}}{r}\right) ^{-2\beta }\left( 1-%
\frac{r_{1}}{r}\right) \left( 1-\frac{r_{2}}{r}\right) ,  \notag \\
R &=&\Lambda \frac{r}{r_{-}}\left( 1-\frac{r_{-}}{r}\right) ^{\beta +\frac{1%
}{2}},\quad \gamma \phi =\log \left( 1-\frac{r_{-}}{r}\right) ,\quad \frac{1%
}{4}-\beta ^{2}=\gamma ^{-2},  \label{f3a}
\end{eqnarray}%
where $\Lambda ^{2}=QL/(|\beta |\sqrt{\frac{3}{2}(36\beta ^{2}-1}\,\,)$,\ $%
r_{1,2}=(r_{-}/2)(6\beta +1\pm \sqrt{36\beta ^{2}-1})$ and $1/6<|\beta |<1/2$%
.

Using Eq. (\ref{potential}), the corresponding potential turns out to be 
\begin{eqnarray}
V_{2}(\gamma ,\phi ) &=&-\frac{2}{L^{2}}e^{-4\gamma \beta \phi }(4\beta
^{2}-1)\left[ -(36\beta ^{2}+1)\cosh (\gamma \phi )+27\beta ^{2}-2-12\beta
\sinh (\gamma \phi )\right]   \notag  \label{f6p} \\
&+&3\beta ^{2}(1+12\beta ^{2})\cosh (2\gamma \phi )+24\beta ^{3}\sinh
(2\gamma \phi ).
\end{eqnarray}

Eqs. (\ref{f3a}) represent a one-parameter (the charge) family of
asymptotically AdS solutions of the model (\ref{f6p}) for a fixed value of
the mass (or temperature). As pointed out in Sect.\ \ref{sect:nohair}, in
the charged case the $C_{1}=0$ solution does not necessarily correspond to
extremal $T=0$ black brane solutions. Moreover, in this case statement $1)$
of the no-hair theorem of the previous section is not useful for guessing
about the existence of a full two-parameter (charge $Q$ and mass $M$) family
of hairy black brane solutions.

The potential (\ref{f6p}) is invariant under the duality symmetry (\ref%
{duality}). The dual solutions are easily obtained using Eq. (\ref{duality})
into Eq. (\ref{f3a}).

The geometrical and thermal properties of the solution (\ref{f3a}) depend on
the value of $\beta$. In the parameter region $-1/2<\beta<-1/6$ we have $%
r_{1},r_{2} < r_{-}$. Because $r_{-}$ is the origin of the radial coordinate 
$r$, there are no horizons and the solution is an extremal $T=0$ solution.
For $1/6< \beta <1/2$ we get $r_{1}>r_{-}$ and we have an horizon. Because $%
r_{1}$ is just a simple (not double) root of $U$, the solution does not
represent an extremal $T=0$ solution.

Solution (\ref{f3a}) for $-1/2<\beta<-1/6$ interpolates between an AdS
vacuum at $r=\infty$ and a conformal Lifshitz solution (\ref{e11a}) with $%
C_{1}=C_{2}=0$ in the near-horizon limit $r\sim r_{-}$. In fact, shifting $%
r\to r+r_{-}$ and expanding near $r=0$, Eq. (\ref{f3a}) becomes 
\begin{equation}  \label{f6c}
\gamma\phi= \ln \frac{r}{r_{-}},\quad U=B\left(\frac{r}{r_{-}}%
\right)^{-4\beta},\quad R=\frac{r_{-}}{L} \left(\frac{r}{r_{-}}%
\right)^{\beta+\frac{1}{2}},
\end{equation}
where $B$ is a constant depending on $\beta$ and $r_{-}$. Solution (\ref{f6c}%
) has the conformal Lifshitz form (\ref{e11a}). On the other hand for $1/6<
\beta <1/2$ we have $U\sim r$ and $R\sim$ const., which seems to indicate
that in this case the solution has to interpreted as an extremal $T\neq 0$
solution.

Analogously to the $Q=0$ case, one can also write down near-extremal,
approximate solutions with an horizon (black brane): 
\begin{equation}  \label{f6d}
\gamma\phi= \ln \frac{r}{r_{-}} ,\quad U=B\left(\frac{r}{r_{-}}%
\right)^{-4\beta}-\mu \left(\frac{r}{r_{-}}\right)^{-2\beta},\quad R=\frac{%
r_{-}}{L} \left(\frac{r}{r_{-}}\right)^{\beta +\frac{1}{2}}
\end{equation}

Also in the charged case, both the $C_{1}=0$ solution (\ref{f6c}) and the
near-extremal solution (\ref{f6d}) are exact solution of an
Einstein-Maxwell-scalar gravity model with an exponential potential given by
the leading term in the $\phi\to\infty$ expansion of the potential (\ref{f6p}%
).

\subsection{Other solutions}

\label{sect:other} In this subsection we present a further example of the
use of our general method for generating exact solutions of (\ref{fed}) with
AdS asymptotic behavior, for $Q=0$, $d=2$ and $\varepsilon=0$.

As ansatz for the scalar field we choose a combination of harmonic functions
in $n+2$ dimensions\footnote{%
We do not limit ourselves to an integer $n$, but we take $n$ real.} 
\begin{equation}
\phi =\frac{\sqrt{2n-1}}{2n}\log \frac{X_{+}}{X_{-}},\qquad X_{\pm }=1\pm
\left( \frac{r_{-}}{r}\right) ^{n},\quad \quad n>\frac{1}{2}.
\end{equation}%
In an asymptotically AdS spacetime, this corresponds to a scalar excitation
near $\phi =0$ of mass 
\begin{equation}
m^{2}=-\frac{n(3-n)}{L^{2}}.  \label{g5}
\end{equation}
The scalar excitation is a tachyon with mass above the BF bound for $1/2<n<3$%
. The PET implies the non existence of black brane solutions for $n\geq 3$.

The Riccati equation (\ref{z1}) is solved by 
\begin{equation}
Y=\frac{r^{2n-1}}{r^{2n}-r_{-}^{2n}}.  \label{rc3}
\end{equation}%
In the uncharged case, $Q=0$, Eq. (\ref{sol1}) with $C_{1}=0$ and $%
C_{2}=1/L^{2}$ gives the solution 
\begin{equation}
U=R^{2}=\frac{r^{2}}{L^{2}}\left[ 1-\left( \frac{r_{-}}{r}\right) ^{2n}%
\right] ^{\frac{1}{n}},\quad \phi =\frac{\sqrt{2n-1}}{2n}\log \frac{%
r^{n}+r_{-}^{n}}{r^{n}-r_{-}^{n}},  \label{h6a}
\end{equation}%
which represents an asymptotically AdS domain-wall solution. As expected
also in this case the solution is a $T=0$ extremal solution. Eq. (\ref%
{potential}) gives the potential : 
\begin{equation}
V(\phi )=-\frac{2}{L^{2}}(\cosh \frac{a\phi }{2})^{2-\frac{2}{n}}\left[
(2-n)\cosh a\phi +(n+1)\right] ,\quad a=\frac{2n}{\sqrt{2n-1}}.  \label{hh2}
\end{equation}%
Notice that this potential is invariant under the duality transformation $%
\phi \rightarrow -\phi $. The potential (\ref{hh2}) smoothly interpolates
between the asymptotical AdS region at $\phi =0$ and the $\phi \rightarrow
\infty $ near-horizon region where the potential has the exponential
behavior 
\begin{equation}
V(\phi )=-\frac{(2-n)}{L^{2}}2^{\frac{2}{n}-2}e^{2\sqrt{2n-1}\,\phi }.
\label{ab1}
\end{equation}%
In the special case $n=2$ the potential takes a very simple form, 
\begin{equation}
V(\phi )=-\frac{6}{L^{2}}\cosh \frac{2\phi }{\sqrt{3}}.  \label{h8}
\end{equation}

Also for the case of the potential (\ref{hh2}) hold the same considerations
concerning the existence of non-extremal black brane solutions as those
discussed in subsection \ref{sect:uncharged} for the potential (\ref{pot1}).

The extremal solution (\ref{h6a}) interpolates between an AdS vacuum at $%
r=\infty$ and a domain wall solution of the form (\ref{e11}) with $C_{1}=0$
near $r=r_{-}$.

This can be easily seen by working in the near-horizon approximation.
Shifting $r\to r+r_{-}$ and expanding near $r=0$, Eq. (\ref{h6a}) becomes at
leading order 
\begin{equation}  \label{h7}
\phi= -\frac{1}{a} \ln \frac{r}{r_{-}},\quad U=R^{2}=D^{2}\left(\frac{r}{%
r_{-}}\right)^{\frac{1}{n}},\quad D=\frac{r_{-}}{L}(2n)^{1/(2n)} .
\end{equation}

Near-extremal approximate solutions with a horizon have the form (\ref{e11})
with $C_{1}\neq0$ and are given by 
\begin{equation}  \label{gm}
U=D^{2}\left(\frac{r}{r_{-}}\right)^{1/n}-\mu \left(\frac{r}{r_{-}}%
\right)^{1-1/n},
\end{equation}
whereas $\phi$ and $R$ are given as in Eq. (\ref{h7}). As expected, solution
(\ref{gm}) is an exact solution of a model with the exponential potential (%
\ref{ab1}).

\section{GENERALIZATION TO $d+2$ DIMENSIONS}

\label{section:ddim}

In this section we generalize the black brane solutions found in the $d=2$
case to $d+2$ dimensions.

\subsection{Domain wall solutions}

We start again from the ansatz (\ref{gamma}). When $d\neq 2$ the Riccati
equation (\ref{z1}) is solved by 
\begin{equation}
Y=\frac{\alpha }{r},\quad \alpha (\alpha -1)=-\frac{2}{d}\ \gamma ^{-2}.
\label{riccati2}
\end{equation}%
It is useful to parametrize $\alpha $ and $\gamma $ as follows: 
\begin{equation}
\alpha =\frac{2}{2d+h^{2}},\text{ \ \ \ }\gamma ^{-1}=\frac{dh}{2+dh^{2}}
\label{pard}
\end{equation}%
Redefining the constant $C_{2}$ in Eqs.\ (\ref{nv}) and (\ref{sol1}), and
rescaling $C_{1}$, the solution takes the form: 
\begin{eqnarray}
U &=&\left( \frac{r}{r_{-}}\right) ^{\frac{4}{2+dh^{2}}}-C_{1}\left( \frac{r%
}{r_{-}}\right) ^{\frac{dh^{2}-2d+2}{2+dh^{2}}},\quad  \label{dw} \\
R &=&\left( \frac{r}{r_{-}}\right) ^{\frac{2}{2+dh^{2}}}, \\
V &=&-\frac{2d[2(d+1)-dh^{2}]}{(2+dh^{2})^{2}r_-^{2}}\ e^{-2h\phi },
\label{dwv}
\end{eqnarray}
As usual, the $r_-$ in the potential can be substituted by the AdS scale $L$%
, using the invariance of the field equations under rescaling of $V$ and $U$.

As in $d=2$, choosing $C_{1}=0$ we obtain the typical domain wall solution $%
U=R^{2}$. For $h^{2}\leq 2/d$, the domain wall solution (\ref{dw}) with $%
C_{1}=0$ has a consistent holographic interpretation and a singularity at $%
r=0$. For $C_{1}\ge 0$ and $h^{2}\leq 2+2/d$ , the solution (\ref{dw}) is
asymptotical to the domain wall solution, and has a horizon at $%
r_{h}=C_{1}^{\;(2+dh^{2})/(2d+2-dh^{2})}r_{-}$.

\subsection{Charged solutions}

The previous solution can be generalized to the case $Q\ne0$. This is the
only charged solution, among those found for $d=2$, that can be computed in
closed form in $d+2$ dimensions.

The ansatz for the scalar field is still given by (\ref{gamma}), whereas the
solution for the Riccati equation is the same as in Eq.\ (\ref{riccati2}).
In the case at hand, it is convenient to choose the following
parametrization for $\alpha$ and $\gamma$: 
\begin{equation*}
\alpha =\frac{h^{2}}{2d+h^{2}},\text{ \ \ }\gamma ^{-1}=\frac{dh}{2d+h^{2}}.
\end{equation*}

Equations (\ref{sol1}), (\ref{riccati2}) and (\ref{potential}) give (after a
rescaling of the integration constant $C_{1}$ and $C_{2}$), 
\begin{eqnarray}
U &=&\frac{2(2d+h^{2})^{2}r_{-}^{2}Q^2}{[2d-(d-1)h^{2}](2d-dh^{2})\Lambda
^{2d}}\left( \frac{r}{r_{-}}\right) ^{2\frac{2d-(d-1)h^{2}}{2d+h^{2}}}\Bigg[%
1-C_{1}\,\left( \frac{r}{r_{-}}\right) ^{-\frac{2d-(d-1)h^{2}}{2d+h^{2}}} 
\notag \\
&&+C_{2}\,\left( \frac{r}{r_{-}}\right) ^{-\frac{4d-2dh^{2}}{2d+h^{2}}}\Bigg]%
,  \notag \\
\bigskip \cr\,R &=&\Lambda \left( \frac{r}{r_{-}}\right) ^{\frac{h^{2}}{%
2d+h^{2}}}, \\
V &=&-\frac{2Q^2}{(2-h^{2})\Lambda^{2d}}\left[ 2e^{-2h\phi }+\frac{%
h^{2}[-2d+(d+1)h^{2}]C_{2}}{[2d-(d-1)h^{2}]}\ e^{-4\phi /h}\right] ,
\label{potenz}
\end{eqnarray}%
In order to make the potential independent from the electric charge, one
must choose the integration constant $\Lambda=(r_-Q)^{1/d}$. As usual, one
can introduce a further length scale $L$ in the potential, by performing a
rescaling of the variables.

For $C_{1}=C_{2}=0$, the solution is conformal to $(d+2)$-dimensional
Lifshitz spacetime. For $h^{2}<2d/(d-1)$, $C_{2}=0$ and $C_{1}>0$ the
solution represents a black brane asymptotical to the conformal Lifshitz
spacetime, with a singularity at $r=0$ and a horizon at $r_{h}=C_{1}^{\;%
\frac{2d+h^{2}}{2d-(d-1)h^{2}}}\,r_-$.

\subsection{Asymptotically AdS solutions}

In order to derive asymptotically AdS$_{d+2}$ solutions of our field
equations (\ref{fed}), we consider again the ansatz (\ref{harmf}), which
expresses the scalar field in terms of a harmonic function $X$ given as in (%
\ref{harmf}).

Near the AdS vacuum the scalar field is tachyonic and has mass: 
\begin{equation*}
m^{2}=-d/L^{2},
\end{equation*}
which is always above the BF bound in $d+2$ dimensions. The Riccati equation
(\ref{z1}) is now solved by: 
\begin{equation}  \label{k8}
R= r \left(1-{\frac{r_-}{r}}\right)^{\beta+\frac{1}{2}},\qquad \frac{1}{4}
-\beta^{2}=\frac{2}{d}\,\gamma^{-2},\qquad -\frac{1}{2}<\beta < \frac{1}{2}.
\end{equation}

We search again for extremal solutions with $C_{1}=0$. Setting $C_{2}=1/L^2$
in (\ref{sol1}) and (\ref{potential}), we obtain the following
asymptotically AdS$_{d+2}$ domain wall solution and the corresponding
potential: 
\begin{eqnarray}
U &=&R^{2}=\frac{r^{2}}{L^{2}}\left( 1-\frac{r_{-}}{r}\right) ^{2\beta +1} ,
\quad \gamma\phi=\ln \left(1-{\frac{r_-}{r}}\right)  \label{d+2} \\
V(\phi ) &=&-\frac{d}{L^{2}}e^{2\gamma \beta \phi }\left\{ \frac{1}{2}%
(d+2)(1-4\beta ^{2})+\frac{1}{2}\left[ 4\beta ^{2}(d+2)+d\right] \cosh
(\gamma \phi )-2\beta (d+1)\sinh (\gamma \phi )\right\} .  \notag
\label{d+2pot} \\
&&
\end{eqnarray}%
One can easily check that the previous potential satisfies, as expected, $%
V(0)=-d(d+1)/L^{2}$ and $V^{\prime }(0)=0$. Notice that the metric part of
the solutions for the generic case (\ref{d+2}) is exactly the same as in the 
$d=2$ case (see Eq. (\ref{f3})). Only the scalar field and the potential are
changed. Also in $d+2$ dimensions the potential (\ref{d+2pot}) is invariant
under the duality transformation (\ref{duality}). Dual solutions are easily
obtained using (\ref{duality}) into Eq. (\ref{d+2}) and (\ref{k8}).

Since the metric functions $U$ and $R$ do not depend on the spacetime
dimension, the near-horizon and near-extremal approximate behavior of $U$
and $R$ is the same as in the $d=2$ case. Thus, the hairy extremal solution (%
\ref{d+2}) always interpolates between an AdS$_{d+2}$ vacuum at $r=\infty $ and a
domain wall solution (\ref{dw}) near $r=r_{-}$.

As in $d=2$, the case $\beta =0$ is particularly simple. The metric part of
the solution is still the same as in $d=2$ and is given by Eq. (\ref{f4b}),
whereas the scalar field and the potential are 
\begin{equation}
\phi =\frac{1}{2}\sqrt{\frac{d}{2}}\log \left( 1-\frac{r_{-}}{r}%
\right),\quad V(\phi ) =-\frac{d}{L^{2}}\left[ \frac{d}{2}\cosh \left( 2%
\sqrt{\frac{2}{d}}\phi \right) +\frac{d+2}{2}\right] .
\end{equation}

For what concerns the existence of nonextremal $C_{1}\neq 0$ solutions in $%
d+2$ dimensions, and the consequences of the no-hair theorem of Sect.\ \ref%
{sect:nohair}, the same considerations as in the $d=2$ case hold.

\subsection{Other solutions}

It is also easy to work out the generalization to $d+2$ dimensions of the
model described in subsection \ref{sect:other}.

We consider the following ansatz for the scalar field: 
\begin{equation}
a\phi =\log \frac{X_{+}}{X_{-}},\quad X_{\pm }=1\pm \left( \frac{r_{-}}{r}%
\right) ^{n},\quad a=\sqrt{\frac{8n^{2}}{d(2n-1)}}\text{ \ \ \ }n>\frac{1}{2}%
.
\end{equation}
Near the AdS$_{d+2}$ vacuum the scalar field is a tachyon with mass 
\begin{equation*}
m^{2}=-\frac{n(d+1-n)}{L^{2}},
\end{equation*}
which is always above the BF bound in $d+2$ dimensions. The PET forbids the
existence of black brane solutions for $n\geqslant d+1$, when the
square-mass of the scalar becomes positive.

The Riccati equation give the same solution (\ref{rc3}) as in the $d=2$
case, the metric function $U$ (with $C_{1}=0$ and $C_{2}=\frac{1}{L^{2}}$)
is given by Eq. (\ref{h6a}), while the potential becomes 
\begin{equation}
V(\phi ) =-\frac{d}{L^{2}}\left( \cosh a\frac{\phi }{2}\right) ^{2-\frac{2}{n%
}}\left[ \left( \frac{d+2}{2}-n\right) \cosh a\phi +\left( n+\frac{d}{2}%
\right) \right].
\end{equation}
Because the metric part of the solution is the same obtained for $d=2$ the
near-horizon, near-extremal approximate solution for $U$ and $R$ are
identical to those obtained in four dimensions.

\section{SPHERICAL AND HYPERBOLIC SOLUTIONS}

\label{sec:spherical} The results of the previous sections can be easily
generalized to the case in which the two-dimensional sections of the
solutions are spherical or hyperbolic. Contrary to the planar case, where it
is dimensionless, the metric function $R$, and hence the integration
constant $\Lambda$ in the solution (\ref{mf}) is now usually taken to have
the physical dimension of a length. Therefore, when $\Lambda$ is not
determined by the field equation, we shall identify it with the AdS length $L
$.

\subsection{Uncharged black hole solutions}

\label{sec:sphericalbh}

We first consider the case of four dimensions ($d=2$). The field equations
are given by (\ref{fed}), with $\varepsilon=\pm1$ and the solutions by (\ref%
{z1})-(\ref{mf}).

The generalization of the black brane solutions of sect.\ \ref%
{sect:unchargedbb} to the case of spherical (or hyperbolic) symmetry is
obtained adopting the ansatz (\ref{gamma}). Substituting the solutions (\ref%
{solution}) with parametrization (\ref{param}), in the general solution of
sect.\ \ref{sect:fieldequation}, after rescaling $C_{1}$ and putting $C_{2}=0
$, the metric functions take the form 
\begin{equation}
U=\frac{(1+h^{2})\varepsilon r_{-}^{2}}{(1-h^{2})L^{2}} \left( {\frac{r}{%
r_{-}}}\right) ^{\frac{2h^{2}}{1+h^{2}}}\left(1-{\frac{C_{1}r_-}{r}} \right)
,\qquad R=L \left( {\frac{r}{r_{-}}}\right) ^{\frac{1}{1+h^{2}}},
\end{equation}%
with potential 
\begin{equation}
V=\frac{2h^{2}\varepsilon }{(h^{2}-1)L^{2}}\ e^{-2\phi /h},
\end{equation}%
having a simple exponential form, as in the planar case. Notice that we have
identified the integration constant $\Lambda$ with the the AdS length $L$.

If $\varepsilon =1$ and $h^{2}<1$, the solutions represent spherically
symmetric black holes with conformal Lifshitz asymptotics, exhibiting a
singularity at $r=0$, shielded by a horizon at \hbox{$r_h=C_1r_-$}.
Solutions exist also for $\varepsilon =-1$ and $h^{2}>1$, they are black
holes with conformal Lifshitz asymptotics and horizons with hyperbolic
topology.

\subsection{Charged black hole solutions}

We now try to extend the previous solutions to the case of nonvanishing
electric charge, generalizing those of sect.\ \ref{sect:chargedbb}. With the
parametrization (\ref{paramc}), the solution reads, after a redefinition of
the constants $C_{1}$ and $C_{2}$, 
\begin{eqnarray}
U&=&\frac{(4+h^{2})r_{-}^{2}}{4-h^{2}}\ \bigg[-\frac{\varepsilon} {\Lambda
^{2}} \left({\frac{r}{r_{-}}}\right) ^{\frac{8}{4+h^{2}}}+\frac{%
(4+h^{2})Q^{2}} {(2-h^{2})\Lambda ^{4}}\left( {\frac{r}{r_{-}}}\right) ^{2%
\frac{4-h^{2}}{4+h^{2}}}+C_{2}\left( {\frac{r}{r_{-}}}\right) ^{\frac{2h^{2}%
}{4+h^{2}}}  \notag \\
&&-C_{1}\left( {\frac{r}{r_{-}}}\right) ^{\frac{4-h^{2}}{4+h^{2}}}\bigg] %
,\qquad\qquad R=\Lambda \left( {\frac{r}{r_{-}}}\right) ^ {\frac{h^{2}}{%
4+h^{2}}},
\end{eqnarray}%
with 
\begin{equation}
V=-\frac{4Q^{2}}{(2-h^{2})\Lambda ^{4}}\ e^{-2h\phi }+\frac{8\,\varepsilon }{%
(4-h^{2})\Lambda ^{2}}\ e^{-h\phi }+\frac{2h^{2}(4-3h^{2})C_{2}}{%
(4+h^{2})(4-h^2)}\ e^{-4\phi /h}.
\end{equation}

Contrary to the planar case, if $\varepsilon \neq 0$, one cannot eliminate
from the potential the dependence on $Q$ by a suitable choice of the
integration constants: one ought in fact to impose $\Lambda ^{2}=Q=1$.
However, the solution with $Q=0$, $C_2\ne0$ may still have interest. For $%
\varepsilon>0$, $h^2>4$ and $C_{1}>0$, such solution represents a black hole
with domain wall asymptotical behavior, a singularity at $r=0$ and one or
two horizons, depending on the value of $C_2$. The asymptotic behavior is
dictated by the $C_{2}$ term. The potential is the sum of two exponential.

\subsection{Asymptotically AdS solutions}

In this section, we wish to generalize the asymptotically anti-de Sitter
solutions obtained using the ansatz (\ref{harmf}), in the case $%
\varepsilon\ne0$. The solution (\ref{f2}) for the radial function still
holds, while, after the usual rescaling of $C_2$, the metric function $U$
becomes in the special case $C_1=0$, corresponding to an extremal black hole,

\begin{equation}
U=-\frac{\varepsilon r^{2}}{2\beta(1+4\beta )L^{2}}\,X^{-2\beta } \left[%
1-(1+4\beta)\frac{r_{-}}{r}\right] +\frac{C_{2}r^2}{L^{2}}\,X^{2\beta +1},
\label{aads}
\end{equation}
where we have rescaled $C_2$, set $L=\Lambda r_-$, and $-1/4<\beta <0$. The
potential is then 
\begin{equation}  \label{poten1}
V(\phi )=-\frac{\varepsilon }{2\beta (1+4\beta )}V_{1}(-\gamma ,\phi
)+C_{2}V_{1}(\gamma ,\phi ).
\end{equation}%
where $V_{1}(\gamma ,\phi )$ is given by Eq.\ (\ref{pot1}). The metric is
singular at $r=0$, while, when $C_{2}\neq 0$, in general the solution is a
black hole, whose horizon structure cannot be determined analytically. As in
the planar case, solutions with $C_{1}\neq 0$ exist, but it is not possible
to eliminate $C_{1}$ from the potential. Hence if a family of black hole
solutions exists for the potential (\ref{poten1}), it must be determined
numerically.

An interesting property of the potential (\ref{poten1}) is the symmetry
between its two terms for $\phi \rightarrow -\phi $. In particular, choosing 
$C_{2}=\varepsilon /{2\beta (1+4\beta )L^{2}}$, the potential becomes 
\begin{eqnarray*}
V(\phi ) &=&-\frac{2\varepsilon }{\beta (1+4\beta )L^{2}}\{(\beta +\frac{1}{2%
})(4\beta +1)\sinh [2(\beta -\frac{1}{2})\gamma \phi ]-(8\beta ^{2}-2)\sinh
[2\beta \gamma \phi ] \\
&&+(\beta -\frac{1}{2})(4\beta -1)\sinh [(2\beta +1)\gamma \phi ]\}.
\end{eqnarray*}

The most interesting case is $C_{2}=0$ and $\varepsilon =1$. With this
assumption, using the invariance of the field equations under the rescaling $%
R\rightarrow \frac{1}{\lambda }R,\,U\rightarrow \lambda ^{2}U,\,V\rightarrow
\lambda ^{2}V$ and changing the sign of $\gamma $ in Eq. (\ref{harmf}), the
potential (\ref{poten1}) can be brought into the form $V=V_{1}(\gamma ,\phi )
$ where, as usual, $V_{1}(\gamma ,\phi )$ is given by Eq. (\ref{pot1}).
Solution (\ref{aads}) becomes 
\begin{equation}
U=-\frac{r^{2}}{L^{2}}X^{-2\beta }\left[ 1-(1+4\beta )\frac{r_{-}}{r}\right]
,\quad R=\frac{L\,r}{r_{-}\sqrt{-2\beta (1+4\beta )}}X^{\beta +\frac{1}{2}%
},\quad \gamma \phi =-\log X.  \label{aads1}
\end{equation}%
In the range of definition, $-1/4<\beta <0$, the solution has no horizon.
However, as we will see when we consider the near-horizon, near extremal
solution, it cannot be considered an extremal black hole.

We conclude by observing that a solution can be found also in the particular
case $\beta =0$, $\gamma =2$. In this case, 
\begin{equation*}
\phi =\frac{1}{2}\log X,\text{ \ \ }R=\Lambda rX^{1/2},
\end{equation*}%
and then 
\begin{eqnarray*}
U &=&\frac{2\varepsilon r(r-r_{-})}{L^{2}}\left[ \frac{r_{-}}{r-r_{-}}+\log
\left( \frac{r-r_{-}}{r}\right) \right] , \\
V &=&-\frac{4\varepsilon }{L^{2}}[4\phi +2\phi \cosh (2\phi )-3\sinh (2\phi
)].
\end{eqnarray*}

\subsection{Charged asymptotically AdS solutions}

We consider now the solutions of the previous section with $Q\neq 0$, but $%
C_{2}=0$. The only change is in the function $U$, 
\begin{eqnarray}
U &=&-\frac{\varepsilon r^{2}}{2\beta (1+4\beta )L^{2}}X^{-2\beta }\left[
1-(1+4\beta )\frac{r_{-}}{r}\right]   \label{equation4} \\
&&+\frac{8\mu r^{2}}{3(1-36\beta ^{2})}X^{-4\beta }\left[ 1+(1+6\beta )\frac{%
r_{-}}{r}+3\beta (1+6\beta )\frac{r_{-}^{2}}{r^{2}}\right] ,  \notag
\end{eqnarray}%
and in the potential 
\begin{equation}
V(\phi )=-\frac{\varepsilon }{2\beta (1+4\beta )}V_{1}(-\gamma ,\phi )-\frac{%
2\mu L^{2}}{3\beta (1+4\beta )}V_{2}(\gamma ,\phi ),  \notag \\
\end{equation}%
where we have defined $L=\Lambda r_{-}$, $\mu =\frac{Q^{2}}{r_{-}^{3}\Lambda
^{4}}$ and $V_{1}(\gamma ,\phi )$ and $V_{2}(\gamma ,\phi )$ are given
respectively by Eqs. (\ref{pot1}) and (\ref{f6p}). Therefore $\mu $, $L$ and 
$\beta $ are parameters of the action, $Q$ is a free parameter and $%
r_{-}=\mu L^{4}/Q^{2}$. Multiple horizons may occur, but cannot be
determined analytically for generic $\beta $.

\subsection{Spherical solutions generated from the planar ones}

For $d=2$, spherical solutions can simply be generated from the planar ones
just by exploiting the fact that the field equations (\ref{fed}) are linear
in the metric function $U$. This method permits to find spherical solutions
for a given form of the potential. This fact may be very useful when one
wants to compare planar and spherical solutions of the same model or when
the method described in the previous subsection gives a singular result
(e.g. $\beta=0$ in Eq.\ (\ref{aads})).

Indicating with $U_{0},\,R_{0}=rX^{\beta +1/2},\phi _{0}=\gamma ^{-1}\ln X$,
where $X$ is the harmonic function (\ref{harmf}), a solution of the field
equations (\ref{fed}) for $\varepsilon =0$ and $d=2$, it follows from the
linearity in $U$ of the field equations that a solution of (\ref{fed}) for $%
\varepsilon =\pm 1$ and $d=2$ is given by 
\begin{equation}
U=U_{0}+\varepsilon X^{-2\beta },\quad R=R_{0},\quad \phi =\phi _{0}.
\label{gen1}
\end{equation}

This method can be used to generate $\varepsilon =\pm 1$ solutions for the
potentials (\ref{pot1}) and (\ref{f6p}) from the planar solutions
respectively given by (\ref{f3}) and (\ref{f3a}). In the uncharged case, i.e
for the potential (\ref{pot1}) we have 
\begin{equation}
U=X^{2\beta +1}\frac{r^{2}}{L^{2}}+\varepsilon X^{-2\beta },\quad
R=rX^{\beta +\frac{1}{2}},\quad \gamma \phi =\ln X,\quad X=1-\frac{r_{-}}{r}%
,\quad \frac{1}{4}-\beta ^{2}=\gamma ^{-2}.  \label{sol45}
\end{equation}%
Notice that, differently from Eqs. (\ref{aads1}) this solution holds in the
full range $-1/2<\beta <1/2$ of the parameter $\beta $. For $\beta =0$ we
get the spherical extremal solution of the model (\ref{gaugedsu}): 
\begin{equation}
U=\frac{r^{2}}{L^{2}}-\frac{rr_{-}}{L^{2}}+\varepsilon ,\quad
R=rX^{1/2},\quad \phi =\frac{1}{2}\ln X.  \label{sol46}
\end{equation}%
One can easily check that  solution (\ref{sol45})is an extremal $T=0$ solution, in
fact the zeros of $U$ are always behind the origin of the radial coordinate
at $r=r_{-}$.

Eqs. (\ref{aads1}) and (\ref{sol45}) are solutions of the same model with
potential $V_{1}(\gamma,\phi)$ given by (\ref{pot1}). Because $%
V_{1}(\gamma,\phi)$ is invariant under the duality transformation (\ref%
{duality}), one can generate from (\ref{aads1}), (\ref{sol45}) two other
solutions of the same model just by reversing the sign of $\beta$ and $\phi$.

\subsection{Near-horizon, near-extremal solution with spherical topology}

Let us now consider the near-horizon approximation of the solutions (\ref%
{aads1}) and (\ref{sol45}). One can obtain the near-horizon, near-extremal
solution by first shifting $r\rightarrow r+r_{-}$ and expanding near $r=0$.
Then one solves the field equations perturbatively near $r=0$ using the the
extremal, near-horizon solution as zero-th order approximation.

In the case of solution (\ref{aads1}) this procedure gives the solution 
\begin{eqnarray}
U &=&\left( \frac{r_{-}}{L}\right) ^{2}x^{-2\beta }\left[ (1-4\beta
(1+2\beta ))x-C\right] ,  \notag  \label{nex1} \\
R &=&\frac{Lx^{\beta +\frac{1}{2}}}{\sqrt{-2\beta (1+4\beta )}}\left[ 1-%
\frac{2}{C}\beta (2\beta -1)x\right] ,\quad \gamma \phi =-\ln x+\frac{4\beta 
}{C}x,
\end{eqnarray}%
where $C$ is an integration constant and $x=r/r_{-}$. For $C>0$ the
solutions describe black holes with a regular horizon at $%
x=x_{h}=C/(1-4\beta (1+2\beta ))$ with $R(x_{h})\neq 0$ and a singularity at 
$r=0$. The Hawking temperature of the horizon is $T=(1/4\pi
)(r_{-}^{2}/L^{2})x_{h}{}^{-2\beta }$. The solution (\ref{nex1}) is singular
for $C=0$, although $T\rightarrow 0$ as $C\rightarrow 0$. For $C<0$ we have
solutions with no horizon and in particular for $C=C_{0}=4\beta $ we get the
solution (\ref{aads1}) in the near-horizon approximation.

Because solution (\ref{nex1}) has a singularity at $r=0$ (corresponding to
the singularity of (\ref{aads1}) at $r=r_{-}$), one should reject solutions
with $C<0$ as unphysical. This gives a strong hint about the nature of our
solution (\ref{aads1}): it is an isolated solution disconnected from the
continuous part of the black hole spectrum at $C>0$ by solutions with naked
singularities.

In the case of solution (\ref{sol45}), near-horizon, near-extremal solutions
are given by 
\begin{eqnarray}
U &=&\frac{r_{-}^{2}}{L^{2}}\left( x^{2\beta +1}-Cx^{-2\beta }\right)
+2\beta x^{-2\beta +1},  \notag  \label{nex2} \\
R &=&r_{-}x^{\beta +\frac{1}{2}}\left( 1+\frac{L^{2}}{Cr_{-}^{2}}(\beta -%
\frac{1}{2})x\right) ,\quad \gamma \phi =\ln x+\frac{L^{2}}{Cr_{-}^{2}}x,
\end{eqnarray}%
where $C$ is an integration constant. For $C>0$ the solutions have a regular
horizon, which at leading order is located at $x=x_{h}=C^{1/(4\beta +1)}$
when $C$ is small, $C<\left( r_{-}^{2}/(L^{2}(1-2\beta ))\right) ^{-(4\beta
+1)/(4\beta )}$. This is consistent with a near-extremal approximation. The
Hawking temperature of the horizon is $T=(1/4\pi )(r_{-}^{2}/L^{2})(4\beta
+1)x_{h}{}^{2\beta }$. The solution (\ref{nex2}) is singular for $C=0$. For $%
C<0$ we have solutions with no horizon and, in particular, for $%
C=C_{0}=L^{2}/r_{-}^{2}$ we get solution (\ref{sol45}) in the near-horizon
approximation.

Also solution (\ref{nex2}) has singularity at $r=0$, which corresponds to
the singularity of (\ref{sol45}) at $r=r_{-}$. Hence, solutions with $C<0$
have naked singularities and solution (\ref{sol45}) is disconnected from the
continuous part of the black hole spectrum at $C>0$.

\section{CONCLUSIONS}

\label{sect:concl}

In this paper, we have presented a general method for finding static,
 radially symmetric, analytic solutions of Einstein and Einstein-Maxwell
gravity minimally coupled to a scalar field. Rather than assuming a
particular form of the scalar self-interaction potential, our method starts
from an ansatz for the scalar field profile and determines, together with
the metric functions, the corresponding form of the potential. For this
reason it is particularly suitable for applications to the AdS/CFT
correspondence. We have investigated in detail two related applications of
our method.

We have first derived a new no-hair theorem about the existence of black
hole solutions of Einstein gravity with scalar hair. As a second
application, we have derived broad classes of exact analytic hairy solutions
of Einstein and Einstein-Maxwell gravity minimally coupled to a scalar.
These solutions have been derived using rather general and simple ans\"atze
for the scalar (in terms of harmonic and logarithmic functions). They cover
many different situations: four or higher dimensions; solutions with planar,
spherical or hyperbolic horizon topology; solutions with AdS, domain-wall
and conformal Lifshitz asymptotics; solutions interpolating between an AdS
spacetime in the asymptotic region and domain-wall or conformally Lifshitz
behaviour in the near-horizon region. Also the class of potentials for the
scalar field characterizing these models are broad ranging from the simple
exponential potential --known to give rise in many situations to domain wall
and Lifshitz solutions \cite%
{Cadoni:2009xm,Charmousis:2009xr,Goldstein:2009cv,Perlmutter:2010qu}-- to
more general forms, that contain as a particular case $\mathcal{N}=8$ gauged
supergravity in 4D truncated to the $U(1)$ sector.

Our investigation has shown that Einstein gravity minimally coupled to a
scalar field has a rich spectrum of solutions with non trivial scalar hair
and AdS asymptotics, which may play an important role in applications of the
AdS/CFT correspondence to condensed matter and strongly coupled QFTs.

Our approach has a main drawback. In some situations it does not allow to
find a full one-parameter family of black holes, i.e.\ the full spectrum of
solutions for different temperatures, but only 'extremal' $T=0$ solutions.
Moreover, what we have called extremal solutions always present a curvature
singularity at $r=0$.

Although our method always allows to find one-parameter families of
near-horizon near-extremal solutions, interpolating solutions with AdS
asymptotics can be found only in the extremal case. Moreover, in many
situations, it is not even clear whether or not such solutions exist. This
is a particularly important question in the cases in which the exact
solution interpolates between the AdS spacetime and a near-horizon domain
wall or Lifshitz spacetime. In the spherical case we have found strong
evidence that our exact solutions represent isolated solutions disconnected
from the continuous part of the spectrum. A final answer to these question
involves numerical computation. We will address these and other issues
concerning the solutions found in this work -- in particular a detailed
discussion of the thermodynamics and of the causal structure of the
solutions and a precise characterization of the spacetime singularities --
in a subsequent paper.

\begin{acknowledgements}
We thank M. Melis and P. Pani for discussions and illuminating 
comments.
This work was partially supported by a grant funded by P.O.R. SARDEGNA 
F.S.E. 2007-2013.
\end{acknowledgements}
\bibliography{exact}

\end{document}